\documentclass[draft]{agujournal2019}
\usepackage{url} 
\usepackage{lineno}
\usepackage[inline]{trackchanges} 
\usepackage{soul}
\usepackage{xcolor}
\usepackage{placeins}
\usepackage[nomarkers, nolists,figuresonly]{endfloat}

\draftfalse

\journalname{JGR: Planets}

\begin{document}

\title{Diversity of new Martian crater clusters informs meteoroid atmospheric interactions}

\authors{T. Neidhart\affil{*}\affil{1}, E. K. Sansom\affil{*}\affil{1}, K. Miljkovi\'c\affil{1}, G. S. Collins\affil{2}, J. Eschenfelder\affil{2}, I. J. Daubar\affil{3}}

\affiliation{*}{Equal contributions}
\affiliation{1}{School of Earth and Planetary Sciences, Space Science and Technology Centre, Curtin University, Perth, Australia}
\affiliation{2}{Department of Earth Science and Engineering, Imperial College, London, SW7 2AZ, UK}
\affiliation{3}{Earth, Environmental and Planetary Sciences, Brown University, Providence, RI, USA}

\correspondingauthor{Eleanor K. Sansom}{eleanor.sansom@curtin.edu.au}

\begin{keypoints}
\item Crater cluster properties inform our understanding of meteoroid fragmentation 
\item Atmospheric influence reflected in crater cluster properties varies with elevation
\item Size-frequency crater distribution in clusters suggests larger impactors are weaker in bulk strength
\end{keypoints}

\begin{abstract}
We investigated 634 crater clusters on Mars detected between 2007 and 2021, which represent more than half of all impacts discovered in this period. Crater clusters form when meteoroids in the 10 kg to 10 ton mass range break-up in Mars' atmosphere to produce a few to a few hundred fragments that hit the ground. The properties of the clusters can inform our understanding of meteoroid properties and the processes that govern their fragmentation. We mapped individual craters $>1$~m within each cluster and defined a range of cluster properties based on the spatial and size distributions of the craters. The large data set, with over eight times more cluster observations than previous work, provides a more robust statistical investigation of crater cluster parameters and their correlations. Trends in size, dispersion and large crater fraction with elevation support weak atmospheric filtering of material. The diversity in the number of individual craters within a cluster, and their size-frequency distributions, may reflect either a diversity in fragmentation style, fragility or internal particle sizes. 
\end{abstract}

\section*{Plain Language Summary}
Between 2007 and 2021, over a thousand newly formed impact craters have been detected on the surface of Mars. Over half are from bodies that have broken up in the atmosphere, forming clusters of craters. We have mapped all the individual craters in each of these clustered impact sites. By investigating the properties of these sites, such as how many individual craters are there, their dispersion and their size, we can improve our understanding of the impacting bodies and how they break up. We find a wide diversity of cluster patterns, with some elevation dependence. This suggests that, although the atmosphere plays a role in meteoroid break up, there is great variety in the impactor bodies themselves.

\section{Introduction} \label{sec:intro}

Meteoroids can disrupt in planetary atmospheres due to aerodynamic drag and the resulting stresses \cite{Baldwin1971,Passey1980}. Fragments that survive to strike the ground at high speed form a suite of tightly grouped craters, called a crater cluster \cite<e.g. >[]{Artemieva2001,Popova2003,Collins2022}, also known as a crater (strewn) field \cite<e.g. >[]{Passey1980, herrick1994effects}. The height above the surface of initial meteoroid break-up, as well as subsequent processes of separation and further fragmentation, depend on atmospheric density, meteoroid strength and speed \cite{Passey1980}. Hence, the sizes and dispersion of craters within clusters can inform our understanding of the physical properties of meteoroids, as well as the atmospheric entry and fragmentation processes in different planetary atmospheres \cite{Hartmann2018}. For example, studies of strewn fields on Earth have informed understanding of fragment distributions, and constrained the lateral spread from possible lift effects, bow shock interactions or separation from rotating meteoroids \cite{Passey1980}. Similar studies have been done on other planetary bodies, including Venus \cite{herrick1994effects} and Titan  \cite{korycansky2005modeling}.

Several models have been developed to explain fragmentation of an impactor during atmospheric entry. These include models in which the impactor breaks up into a dust cloud and fragments disperse under the differential pressure between the front and back surfaces \cite{Hills1993, Collins2005}, models that assume an object breaking progressively into individual fragments \cite{Passey1980,Artemieva2001, Popova2007,Register2017} and hybrid models that combine both effects \cite{Register2017,Wheeler2017}. On Earth, fireball observations can be used to inform models, though the number of unknown parameters usually requires assumptions be made, or significant computational resources for Monte Carlo style approaches  \cite<e.g.>[]{Sansom2019}. Fireball light curves can be used to identify atmospheric fragmentation events \cite{Borovicka2020}, though the number and size of fragments produced are not directly observable. The properties of crater clusters are highly complementary to fireball studies on Earth, and can provide vital additional constraints to calibrate numerical models of meteoroid fragmentation \cite<e.g.,>[]{collins_meteoroid_2022}. 

Of the 1,203 newly formed impact sites on Mars that have been observed to date, 58\% are classified as crater clusters \cite{Daubar2022}. Mars clusters therefore offer a powerful opportunity to test and refine atmospheric break-up models and constrain meteoroid properties. Previous work by \citeA{Daubar2019} investigated crater clusters on Mars prior to NASA's InSight mission to inform potential seismic energy signatures expected. Here we describe and analyse a dataset of 634 crater clusters that expands on and includes the crater clusters mapped by \cite{Daubar2019}. Entry dynamics and fragmentation strongly depend on atmospheric density and pressure \cite{Ceplecha1998, Collins2005}. As the atmospheric density within 10 km of the surface of Mars is a strong function of surface elevation, the properties of Martian crater clusters may show a correlation with elevation, as has been suggested by both \citeA{Daubar2019} and \citeA{Ivanov2009}. Using this expanded dataset we therefore investigate variations in cluster properties with elevation and the size-frequency distributions of craters within each cluster. The crater cluster mapping results are applied to understanding of: a) atmospheric effects on impactors with variation in elevation of impact sites, b) impactor material fragility, and c) fragmentation mechanics, in support of models for the 
meteoroid fragmentation in the Martian atmosphere and the formation of crater clusters \cite<e.g.,>[]{Collins2022}.

\section{Methods for mapping crater clusters} \label{sec:meth}

Here we mapped 557 crater clusters on Mars that formed between 2007 and 2021, that have since been listed in the most recent crater catalogue \cite{Daubar2022}. The mapping methodology applied was similar to previous works \cite{Daubar2019, Collins2022}. We used HiRISE images \cite{McEwen2007} to map individual craters larger than 1 m in diameter within each of the 557 crater clusters. Together with the 77 previously investigated crater clusters of \citeA{Daubar2019}, this study examined a total of 634 crater clusters from the catalogue of 1203 impact sites (single and clusters) \cite{Daubar2022}. The HiRISE images of the crater clusters used for mapping in this study had a resolution of 0.25 m/pixel, except for 12 crater cluster images that have a lower resolution of 0.5 m/pixel \cite{McEwen2007}. We used ArcMap (ArcGIS) software, with the CraterTools add-on (Figure \ref{fig:1_mapping_ellipse}) to record the geographical location (latitude and longitude) of each crater centre. 

The elevation for each crater cluster was determined from the MOLA 128 ppd basemap \cite{Zuber1992,Smith2001}. Here, we use the elevation for each crater cluster that was determined by collaborators and reported in \citeA{Daubar2022}. The MOLA map used has a horizontal positional accuracy of about 300 m and a vertical positional accuracy of 30 m \cite{Zuber1992}, which is large given the mean cluster footprint of $\sim50\,m$. The elevation for each cluster is therefore given as an average as it is impossible to determine elevations for individual craters within the cluster.

\subsection{Measured and calculated crater cluster parameters}
 
\begin{figure}[!htbp]
\begin{center}
\includegraphics[width=\linewidth]{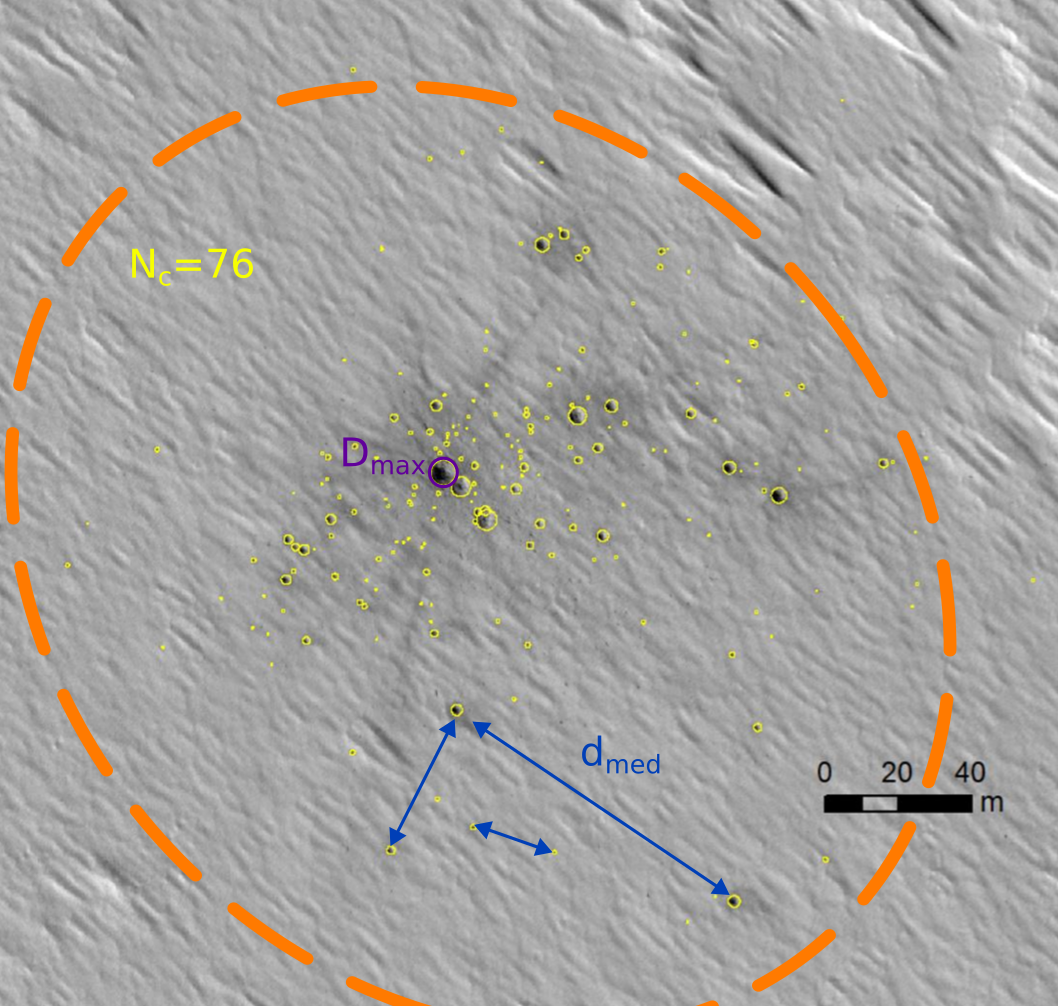}
\caption{Crater cluster ESP\_066902\_1830 showing parameters used to describe clusters in this study: mapped individual craters of the cluster (yellow; with total number mapped as N$_c$), calculated best fit ellipse (orange dashed perimeter), largest crater of the cluster (purple; D$_{max}$) and illustration of separation distances used in dispersion calculation(blue arrows; d$_{med}$). The cluster is located at 2.890$^\circ$N, 254.162$^\circ$E. It consists of 76 craters each with their own blast zones (seen as dark rays here) with an effective diameter of 11.4 m. The diameter of the largest crater in the cluster is 7.1 m, the dispersion of the cluster is 66.7 m, the aspect ratio of the best fit ellipse is 0.75 and it is at an elevation of 5758.8 m. North is up. HiRISE image credit: NASA/JPL/University of Arizona.}
\label{fig:1_mapping_ellipse}
\end{center}
\end{figure}

Several crater cluster parameters were investigated in order to compare the new crater clusters to those previously studied in \citeA{Daubar2019}. These are illustrated in Figure \ref{fig:1_mapping_ellipse} and definitions summarised in Table \ref{tab:properties_description}. Calculated parameters followed the methods from previous works \cite{Daubar2019, Collins2022}.

\begin{table}[h]
\caption{Properties used in this study to describe characteristics of crater clusters.}
\centering
\begin{tabular}{l|l}
\savehline
  Parameter  &  Description  \\\\\savehline
   $N_c$ & Number of craters within a crater cluster $\ge$ 1 m in diameter \\ 
   \savehline
    $D_{max}$ (m) &  Diameter of the largest crater within a crater cluster \\
\savehline
   $D_\mathrm{eff}$ (m) & The effective diameter of an equivalent single crater \\& $D_{\mathrm{eff}}=\sqrt[3]{\sum_iD_i^3}$, where $D_i$ is the diameter of individual craters in cluster \\
  \savehline
   $d_{med}$ (m) & Spatial distribution (or spread) of craters in a cluster calculated as the median\\& of the distances between all crater pair combinations in a cluster. \\
   \savehline
   $e$ & Aspect ratio of the minor to major axis of the best fit ellipse \\
   \, & $e\approx$ 1: circular distribution of craters, \\
   \, &e$<<$ 1: elliptical distribution of craters \\
   \savehline
   F-value & Fraction of craters $>D_{max}/2$ to the total number of craters: $\frac{N(>D_{max}/2)}{N_c}$ \\
   \, & F-value $\approx$ 1: craters comparable in sizes \\
   \, & F-value $<<$ 1: clusters consisting of few larger and many small craters  \\
   \savehline
   DSFD slope &  
   Represents the power law exponential $a$ in $y = C \times x^a$ of the regression \\\,& curve used to fit the Differential Size Frequency Distribution (DSFD) \\\,& of craters in a cluster.\\
   \savehline
      Elevation (m) & Elevation of crater cluster site, from \citeA{Daubar2022}\\
 \savehline
 \end{tabular}
 \label{tab:properties_description}
 \end{table}

\begin{figure}[!htbp]
\begin{center}
\includegraphics[width=\linewidth]{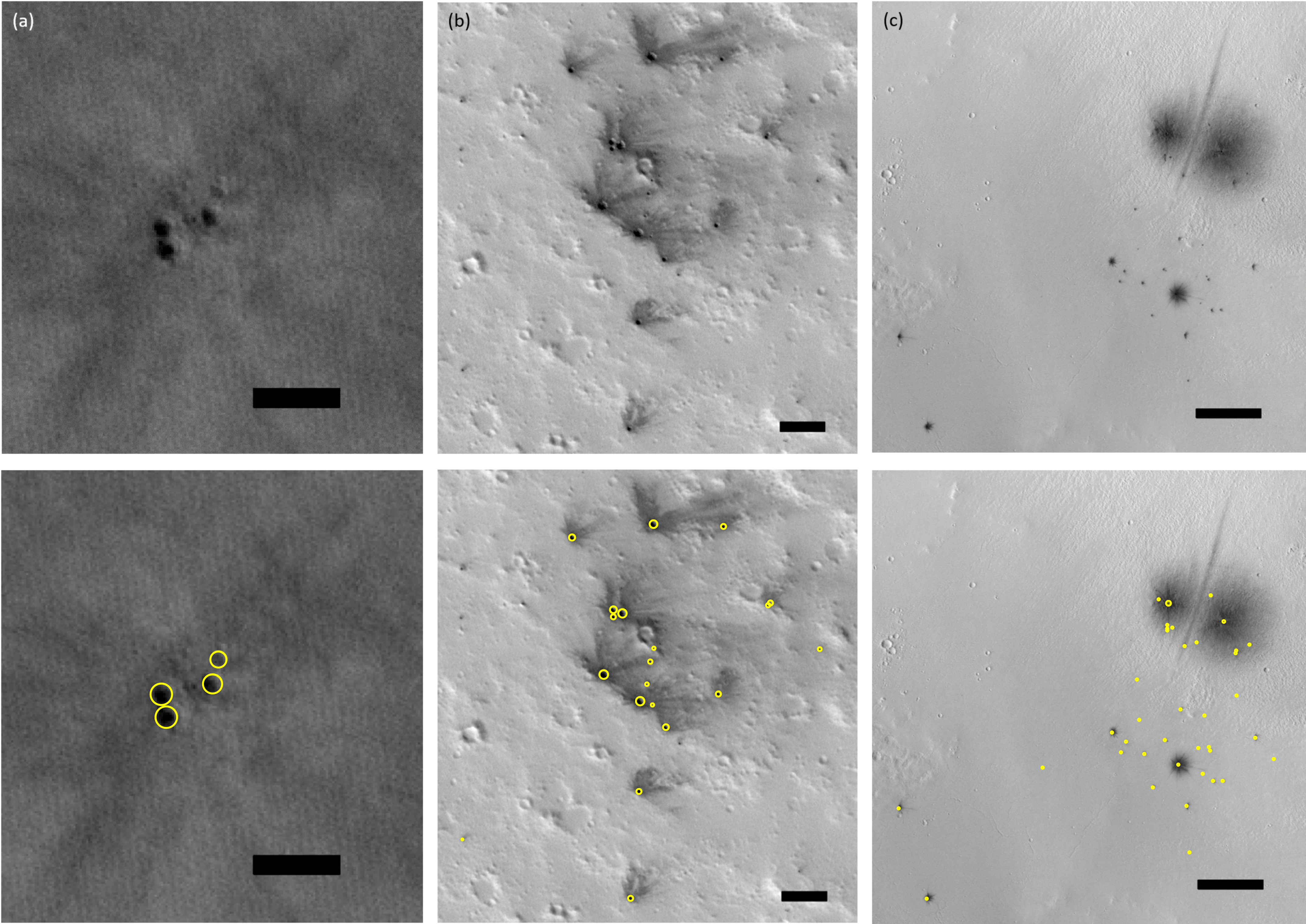}
\caption{Examples of crater clusters having a low (a), median (b) and high dispersion (c) and their marked craters (lower panel). Note that the diameter of crater rim outlines in (c) have been exaggerated for illustrative purposes and are not to scale. (a) Crater cluster ESP\_031912\_1900 located at 9.8865 $^{\circ}$N and 39.0291 $^{\circ}$E having a dispersion of 6.25 m. Scale bar is 10 m. (b) Crater cluster ESP\_030243\_1985 at 18.2744 $^{\circ}$N and 244.9039 $^{\circ}$E with a dispersion of 59.63 m. Scale bar is 20 m. (c) Crater cluster ESP\_011618\_1885 at 8.6248 $^{\circ}$N and 46.8301 $^{\circ}$E  with a dispersion of 254.50 m. Scale bar is 100 m. North is up. HiRISE image credit: NASA/JPL/University of Arizona.}
\label{fig:2_dispersion_examples}
\end{center}
\end{figure}

The total number of individual craters mapped for each cluster is given as $N_c$, with the diameter of the largest member noted as $D_{max}$. The effective diameter ($D_\mathrm{eff}$) provides an estimated diameter of a single crater of equivalent volume. As the diameter of an impact crater can be linked to impactor mass, the $D_\mathrm{eff}$ can inform the total impactor mass distributed across the crater cluster strewn field. The spatial distribution and spread of craters in a cluster is described by the dispersion ($d_{med}$). This is calculated as the median of the distances between all crater pair combinations in a cluster. This definition is consistent with that from \citeA{Collins2022}. Note that this differs from the definition of dispersion in \citeA{Daubar2019} who instead used the standard deviation of distances between all crater pair combinations. The definition from \citeA{Collins2022} was preferred as distances between individual craters in the cluster are not normally distributed. Clusters with high dispersion consist of widely separated craters, whereas those with low dispersion are tightly spaced (see crater cluster examples in Figure \ref{fig:2_dispersion_examples}).
To quantify the shape of the area of the field of individual craters, the best-fit ellipse (red ellipse in Figure \ref{fig:1_mapping_ellipse} is an example) was calculated. The best-fit ellipse is defined as an ellipse with a minimum area which encases the majority of the craters within the cluster. Best-fit ellipses are calculated using a Khachiyan algorithm based solver \cite{Todd2007,Daubar2019,collins_meteoroid_2022}. Individual fits were calculated 80 times using random samples from the distribution of crater locations, via a bootstrapping algorithm \cite{Efron1986}, and the average ellipse is used. This is done to minimise the influence of outliers on the overall shape of the ellipse. The best-fit ellipse is calculated for clusters with more than four craters. Less than four craters would not provide a confident ellipse fit. The aspect ratio (minor axis/major axis) of the best-fit ellipse, $e$, is used to define the shape of a crater cluster: $e \approx$ 1 corresponds to a distribution of craters in a circular pattern within a cluster and $e <<$ 1 corresponds to an elliptical distribution of craters in the cluster. Figure \ref{fig:1_mapping_ellipse} shows a crater cluster shape that is somewhat elliptical (e=0.75).

\begin{figure}[!htbp]
\begin{center}
\includegraphics[width=\linewidth]{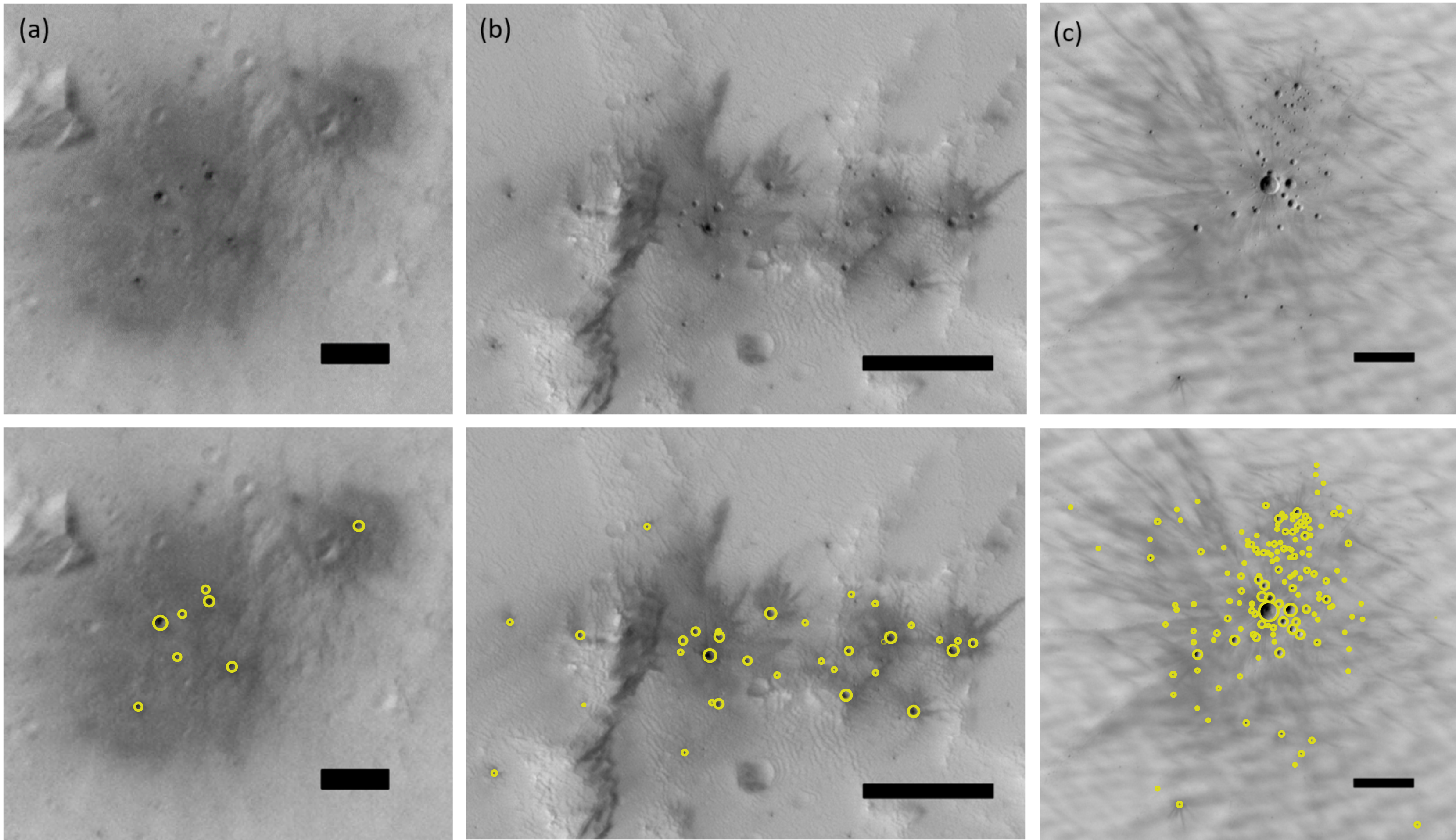}
\caption{Examples of crater clusters having a high (a), median (b) and low F-value (c) and their marked craters (lower panel). (a) Crater cluster ESP\_034319\_1625 located at -17.2769 $^{\circ}$N and 207.3625 $^{\circ}$E having a F-value of 1. Scale bar is 10 m. (b) Crater cluster ESP\_049600\_1815 at 1.322 $^{\circ}$N and 266.804 $^{\circ}$E with a F-value of 0.353. Scale bar is 40 m. (c) Crater cluster PSP\_007036\_1765 at -3.6161 $^{\circ}$N and 234.2425 $^{\circ}$E with a F-value of 0.021. Scale bar is 40 m. North is up. HiRISE image credit: NASA/JPL/University of Arizona.}
\label{fig:3_F_value_examples}
\end{center}
\end{figure}

\begin{figure}[h]
\centering
\includegraphics[width=0.6\textwidth]{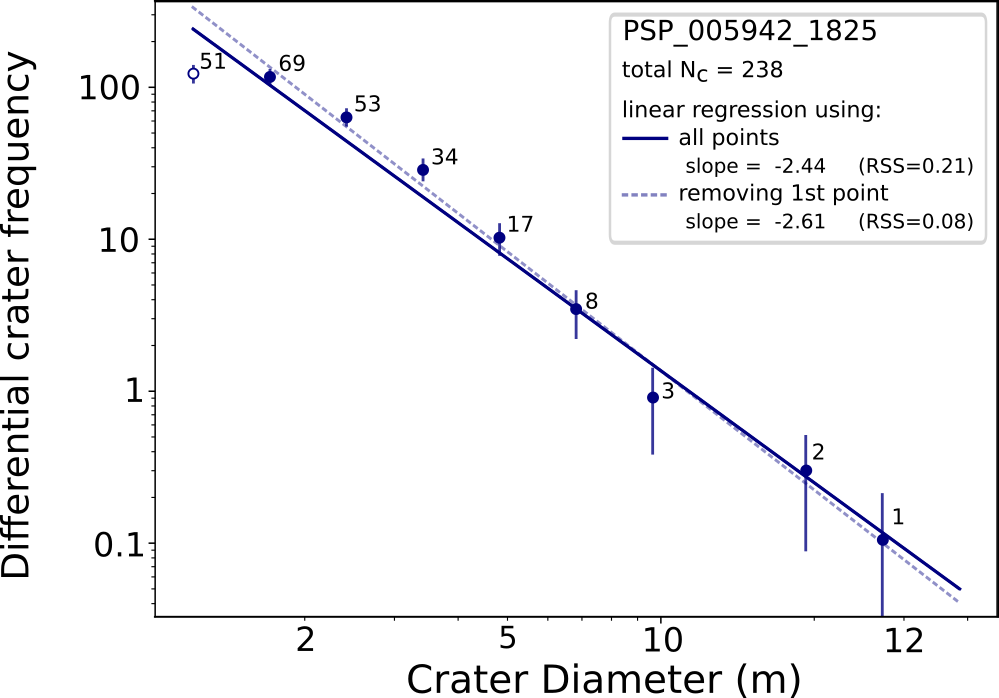}
\caption{An example of fitting a linear regression to the differential size frequency distribution (DSFD) for cluster PSP\_005942\_1825 (solid line, using all points). As a measure of fit, the residual sum of squares (RSS) is also given. Number of craters in each bin (values given above each point) are normalised to the bin width. In this example, the linear regression is also calculated  excluding the first point (dashed line) to show the effect of possible roll over in this smallest diameter bin.}
\label{fig:4_SFD}
\end{figure}

To quantify the population of craters within a crater cluster, the size-frequency distributions were determined. To describe the size-frequency distribution of craters within clusters, we calculate several metrics for each cluster: the diameter of the largest crater ($D_{max}$); the fraction of craters in the cluster
that have a diameter larger than half the maximum crater
diameter (F-value$=\frac{N(>D_{max}/2)}{N_c}$) \cite{Newland2019}; and the slope (in log space) of the differential size-frequency distribution (DSFD) of craters within the cluster. 
The differential size-frequency distribution of craters in a cluster is defined as the number of craters in a given diameter bin, similar to a histogram, but normalised by the bin width \cite{robbins2018revised}; the differential frequency for a crater of diameter $D_x$ is $N(D_x) = n[D_{x1}, D_{x2}] / (D_{x2} - D_{x1}) $. Here we use a bin width of $\sqrt{2}D$, initialising with a diameter ($D$) of 1\,m. 
A linear regression is then calculated in log-log space and the resulting slope, $a$, in $log(y) = a \times log(x) + c$ represents the power-law exponent in $y = C x^a$. A residual sum of squares is calculated as a measure of the fit. See Figure \ref{fig:4_SFD} for an example crater cluster DSFD and linear regression fitting for evaluating slopes. Shallower slopes ($\approx$ 0) indicate clusters having similar numbers of small and large diameter craters per unit area. Steeper slopes ($<-2$) indicate clusters with relatively more small diameter craters per unit area.
The F-value and DSFD slopes were not previously presented in the work of \citeA{Daubar2019}, and so we additionally calculated them for those 77 previously investigated crater clusters. Clusters with $F-value \approx 1$ comprise craters with similar sizes, whereas clusters with $F-value << 1$ consist of a few large diameter craters and many smaller ones (see Figure \ref{fig:3_F_value_examples}).

\subsection{Observational bias}
Craters formed in the last 1-2 decades in dusty areas of Mars typically have a recognisable blast zone, where the albedo is different (usually lower) relative to the pre-existing surface. This is likely due to disturbance of surficial dust from effects of the impact event. In some rare cases, it was difficult to determine which craters belong to a cluster, especially in areas with large blast zones and/or a large number of small, similar-sized fresh-looking background craters. The main criteria used to determine that a crater belongs to a cluster was that the crater must (a) be clearly associated with an albedo change relative to the surroundings (i.e. each individual crater had its own blast zone); and (b) appear fresh, with no aeolian bedforms, rim erosion, or any other signs of modification or age. Obvious secondary craters were also excluded from the crater measurements; these were identified as highly elliptical craters directed away from a primary impact site, small relative to the primary, and sometimes forming rays. 

The confidence in crater diameter measurements depended on the type of the terrain where craters were found, especially in cases where blast zones were either very dark or very bright. Where possible, colour images were used which improved the visibility of the crater rim. Alternatively, stretching of the image brightness or contrast was applied where necessary to better identify the blast zone. For non-circular craters or craters extremely close together crater size measurements were not straightforward. For crater clusters where ice made it impossible to recognize crater rims, surface images that showed the cluster after the ice sublimated were used if available. Any non-circular craters were mapped in a way so that the majority of the crater area was included, and overlapping craters were mapped separately where possible. For craters that show terraces (or benches) within the crater, the outer rim diameter was measured. The reliability of the crater mapping method was tested by the independent mapping of a subset of the data set by a second author. The average difference in the diameter of the largest crater in the cluster measurements and the effective diameter measurements was between 16\% and 11\%, respectively, which is typical for non-expert mappers \cite{Robbins2014}, for craters $\geq$ 1 m. The higher variation in comparison to expert crater mappers which is 5-10\% \cite{Robbins2014} can be explained by the very small crater sizes and number of craters in a cluster which was sometimes hard to determine especially if the cluster consists of many small craters close to 1 m in diameter.

\section{Results}

\begin{figure}[h]
\centering
\includegraphics[width=\textwidth]{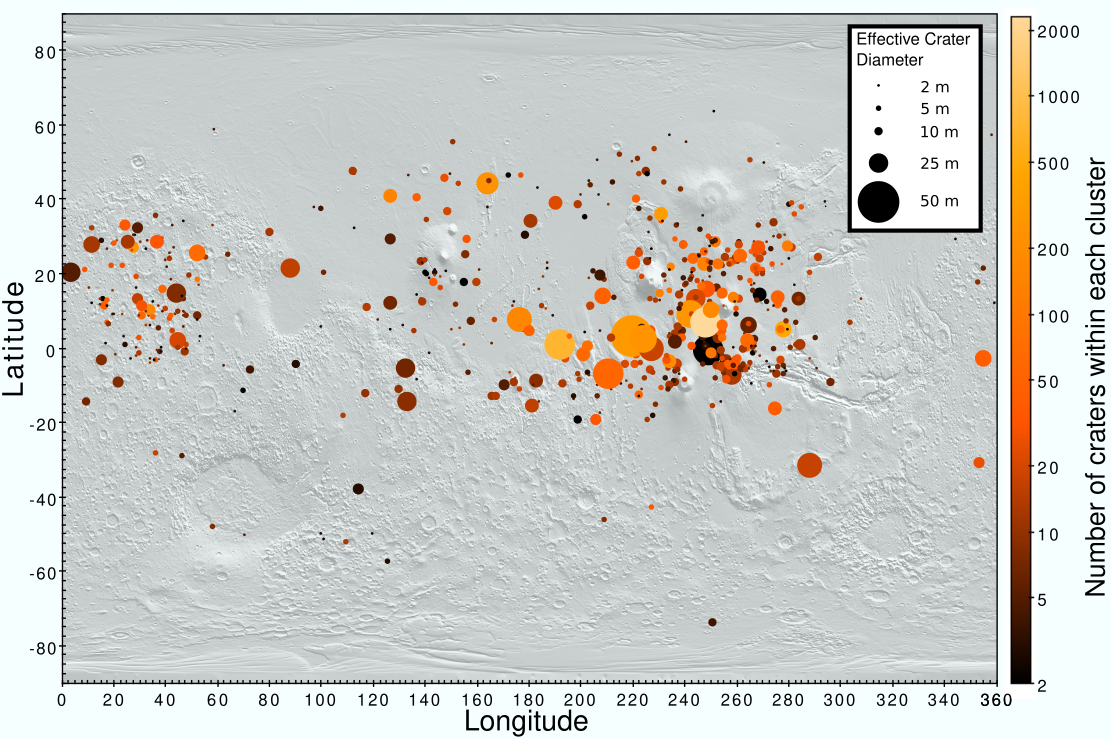}
\caption{Crater clusters on Mars from the catalogue of \cite{Daubar2022} mapped in this study and \citeA{Daubar2019} shown on the topographic shade map of Mars \cite{Neumann2001}. The marker size denotes the cluster effective diameter. Orange colour tones represent the number of craters in a cluster.}
\label{fig:5_map_plain}
\end{figure}

Here we investigate the trends in parameters of 634 crater clusters from the catalog of 1203 fresh impact sites on Mars \cite{Daubar2022}, combining our 557 newly mapped crater clusters with the 77 previously investigated by \citeA{Daubar2019}. Figure \ref{fig:5_map_plain} shows the distribution of all 634 crater clusters on Mars and their relative sizes (both in effective diameter and quantity of individual craters). Our data set represents over 90\% of the total number of crater clusters included in the catalog by \citeA{Daubar2022}; clusters not included were added to the catalog after we had completed portions of our analysis. Furthermore, there are five craters that were originally identified in \citeA{Daubar2022} as single craters, but have been remapped in this study as crater clusters, along with approximately 10 other craters with slightly different size and number of crater measurements. This degree of uncertainty is typical of crater measurements made by different mappers \cite{Robbins2014}. 

\begin{figure}[h]
\noindent\includegraphics[width=\textwidth]{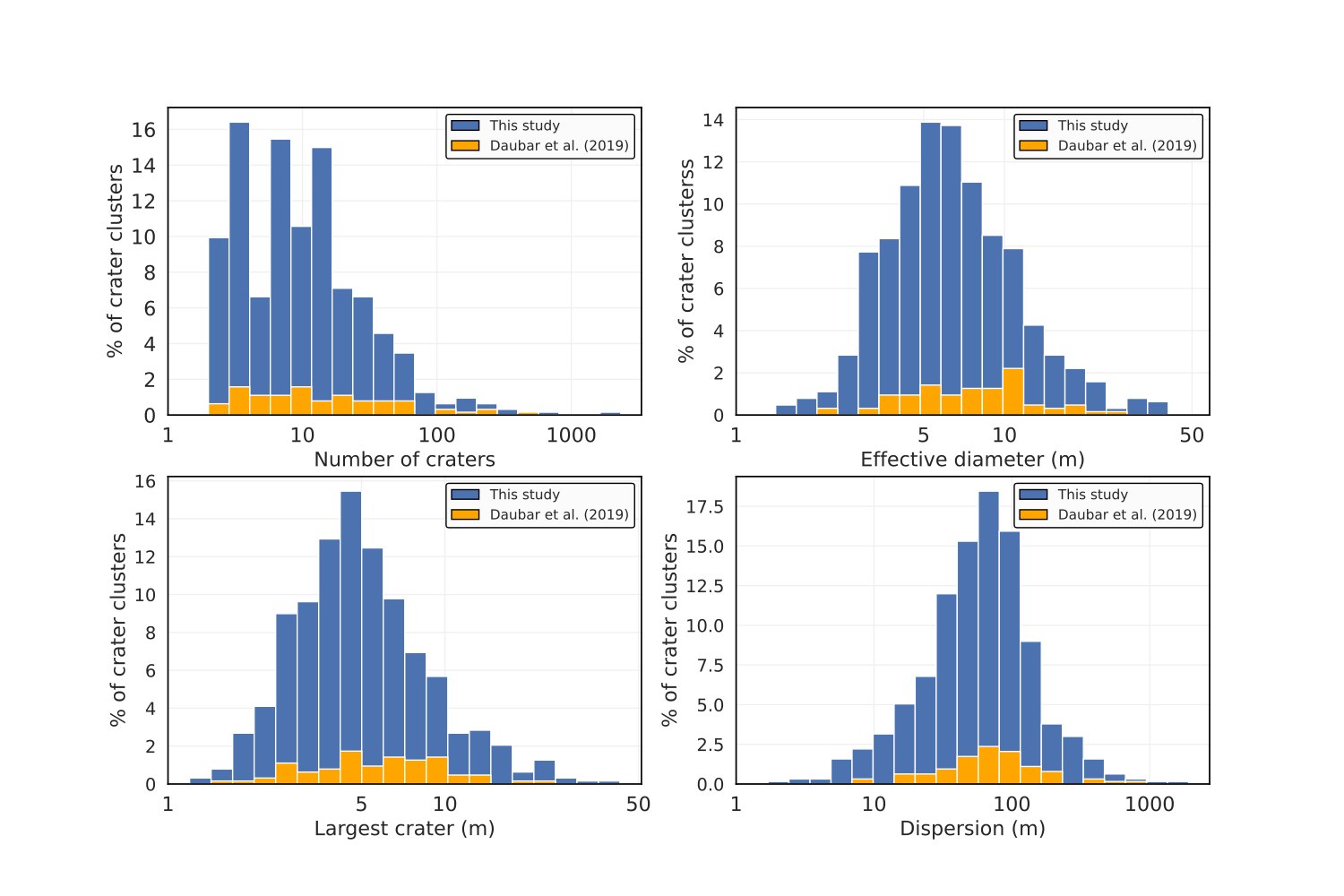}
\caption{Stacked histograms showing the percentage of crater clusters compared to the total number of crater clusters mapped in this work and \citeA{Daubar2019}, as: (a) the number of craters in a cluster, (b) effective diameter, (c) diameter of the largest crater in the cluster and (d) the dispersion (median separation distance between craters). Note that x-axes are shown in logarithmic scale.}
\label{fig:6_histograms}
\end{figure}

Table \ref{tab:properties} summarises the general properties of all 634 observed crater clusters. Values for the 77 clusters from \citeA{Daubar2019}'s previous work are given in brackets for comparison. Figure \ref{fig:6_histograms} summarises the distributions of key crater cluster properties: N$_c$, D$_{eff}$, D$_{max}$ and d$_{med}$. Each distribution was divided into 20 bins in log space, using the domain ranges given in Table \ref{tab:properties}.

\begin{table}[h]
\caption{Measured and calculated crater cluster parameters examined in this study. We report parameters from the total of 634, showing the range, mean values and standard deviation (stdev), compared in parentheses with the 77 clusters studied by \citeA{Daubar2019}.}
\centering
\begin{tabular}{l c l l l}
\hline
  Parameter  & Range & Median & Mean & Standard Deviation  \\
 \hline
   $N_c$ & 2--2334 (2--465) & 9 & 23.8 (34.1) & 103.2 (70.3) \\ 
   $D_\mathrm{eff}$ (m) & 1.4--48.7 (2.1--33.8) & 6.1 & 7.5 (9.1) & 5.5 (5.5) \\
   $D_{max}$ (m) & 1.2--42.9 & 4.6 & 5.9 & 4.5 \\
   $d_{med}$ (m) & 1.7--1914.5 & 59.7 & 89.2 & 133.0 \\
   $e$ & 0.01--0.92 & 0.45 & 0.45 & 0.18 \\
   F-value & 0.004--1 & 0.369 & 0.454 & 0.305 \\
   Elevation (m) & -6724--17914 & 945 & 1130 & 3142 \\
   DSFD slope & -5.6--0 & -1.6 & -1.7 \\
  \hline
\label{tab:properties}
 \end{tabular}
 \end{table}

\subsection{Correlations between crater cluster parameters} \label{subsec:corr}

\begin{figure}[h]
\noindent\includegraphics[width=\textwidth]{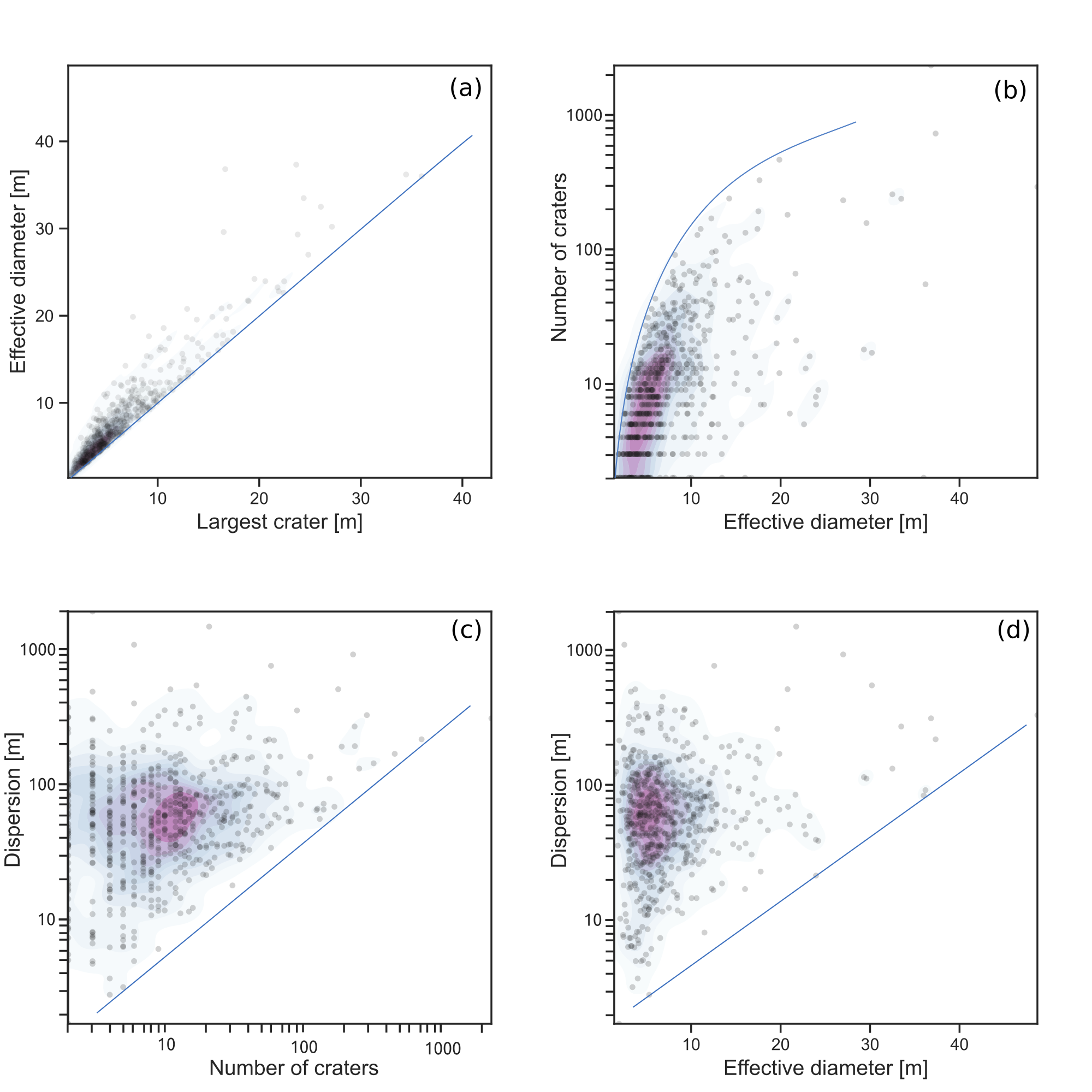}
\caption{Visualising crater cluster properties a) diameter of the largest crater in the cluster and the effective diameter of the cluster, b) number of craters in the cluster for effective diameters, c) dispersion (median crater separation) for number of craters in the cluster, and d) dispersion and effective diameter of the cluster for crater clusters, of this study and \citeA{Daubar2019}. Note that the x-axis in (c) and y-axis in (b, c, d) are shown in logarithmic scale, and that the Kernel Density Estimate (KDE) shading, and approximate data limits (blue lines) are merely to aid in the visualisation of data distributions and boundaries respectively; in (a) this is at 1:1 but for others this line has been added merely for visualisation.}
\label{fig:7_general}
\end{figure}

We examine the trends in crater cluster properties and search for correlations between different characteristics of crater clusters (Figure \ref{fig:7_general}). Kernel Density Estimate (KDE) plots, using the Scott method, allow us to visualise and estimate the multivariate probability density distribution of the cluster characteristics, which are useful for the validation and calibration of numerical fragmentation models \cite{Collins2022}. The commonly used Pearson's correlation is used to measure the strength of the linear relationship between two sets of data \cite{LeeRodgers1988}. 

There is a strong correlation between the effective diameter ($D_{\mathrm{eff}}$) and the largest crater within the cluster (D$_{max}$) (Figure \ref{fig:7_general}a) having a Pearson's correlation coefficient of $\rho$=0.95. The strong correlation implies $D_{max}$ can be used as a proxy for the cluster size. Careful mapping shows that many clusters plot very close to this 1:1 line. $D_{\mathrm{eff}}$ can never be smaller than a cluster's D$_{max}$ by definition, so there is a restricted zone below this 1:1 line. At larger $D_{\mathrm{eff}}$, it appears that the largest crater becomes dominant in the cluster; in other words, the largest crater in the cluster is the main contribution to the effective crater size, the upper limit is $D_{\mathrm{eff}}$ = 2.6 $D_{max}$ and the lower limit is $D_{\mathrm{eff}} \approx D_{max}$. Figure \ref{fig:7_general}b shows that the maximum number of craters in a cluster increases with $D_{\mathrm{eff}}$. 
Figure \ref{fig:7_general}c suggests that the number of craters in a cluster require a minimum dispersion. Those with a small number of individual craters can have a large range of dispersions, but clusters with large number of individual craters tend to have exclusively larger dispersions. 
Figure \ref{fig:7_general}d shows that while clusters with small effective diameters can exhibit a wide range of dispersions, there is potentially a minimum dispersion applied to large clusters. This suggests that larger clusters are more widely spread, though the overall number of observations for these larger clusters could be biasing this result. Clusters with high dispersion and low effective diameter could suggest a highly oblique impact angle, where late fragmentation can still show dispersion due to the incidence angle with the ground. 

\subsection{The size-frequency distribution of craters in clusters}

\subsubsection{Fraction of large craters: F-value}
\begin{figure}[h]
\noindent\includegraphics[width=\textwidth]{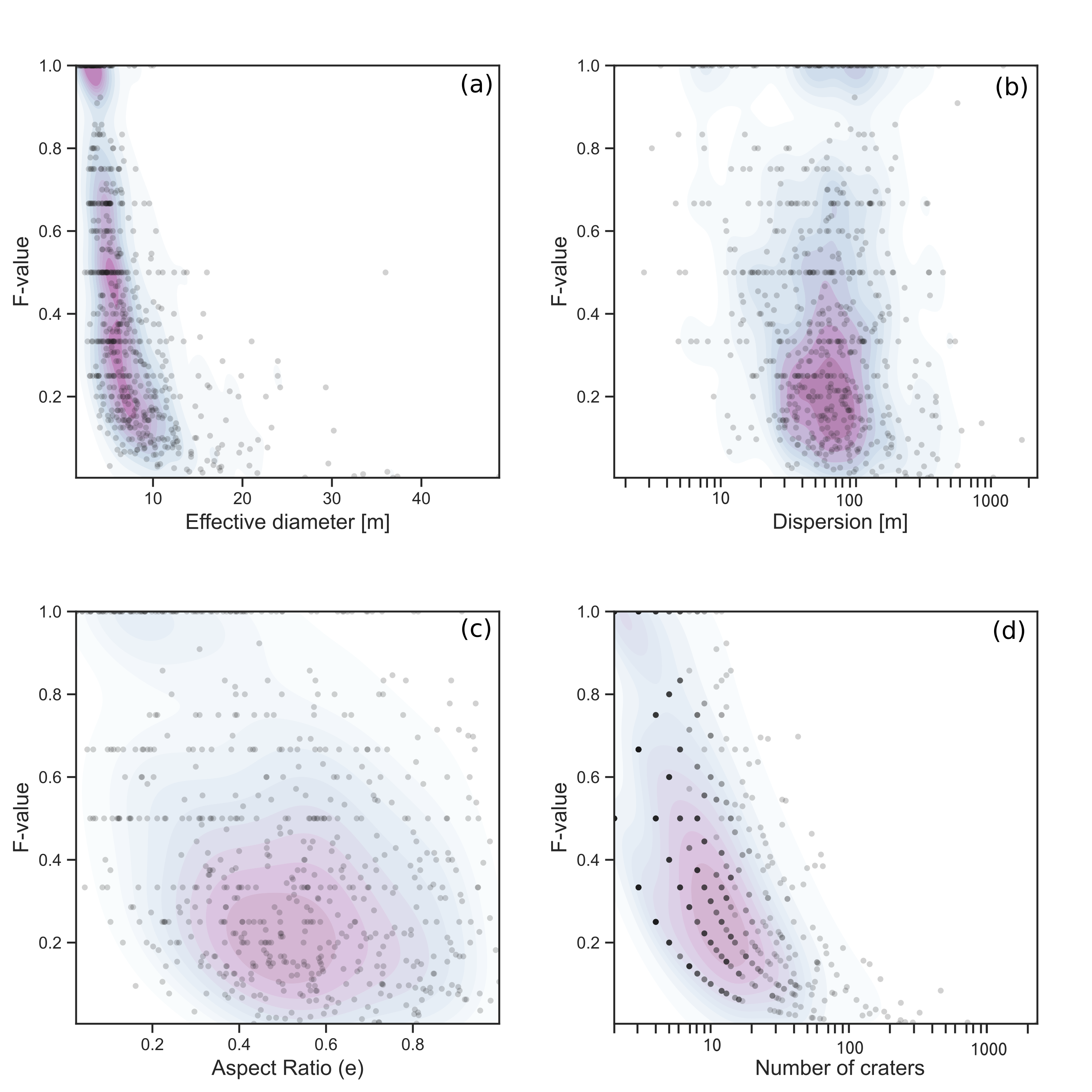}
\caption{Plots showing the F-value for craters with different a) effective diameters, b) dispersion, c) aspect ratios and d) number of craters within a cluster, using data from this study and \citeA{Daubar2019}. Clusters having craters comparable in size have F-value$\approx$1, whereas F-value$<<$ 1 describes clusters consisting of a few relatively large diameter craters and many small ones. Note that the x-axis in (d) is shown in logarithmic scale. Horizontal lines in data are due to similar and common F-values of crater clusters. Note that the Kernel Density Estimate (KDE) shading is merely to aid in the visualisation of data distributions.}
\label{fig:8_f_L}
\end{figure}

Figure \ref{fig:8_f_L} investigates correlations between the F-value and other cluster parameters. Clusters of similar sized diameter craters have F-values $\approx$ 1, whereas F-value $\ll$ 1 describe clusters consisting of a few relatively large craters and many small ones. 

Our observations of F-values showed that clusters with effective diameters up to approximately 10 m (80\% of clusters) have a large variety in relative crater diameters. However, clusters with effective diameters larger than 10 m contain mostly small craters with only a small fraction of large craters (small F-values) (Figure \ref{fig:8_f_L}a). 
There appears to be very little correlation between dispersion and the F-value (Figure \ref{fig:8_f_L}b). Nor is there a  strong correlation between the aspect ratio and F-value (Figure \ref{fig:8_f_L}c). Although, recent modelling has shown that there is a correlation between impact angle and the dispersion of craters within a cluster, particularly for the more elliptical clusters \cite{Collins2022}, here we see a larger range of aspect ratios, which suggests all impact incidences are likely represented in the dataset. Clusters containing a large number of craters (e.g., $N_c$ $>$ 100; 3\% of clusters) have F-values of 0.2 or lower (Figure \ref{fig:8_f_L}d). This means that if a cluster contains a large number of craters, the majority of craters are very small in diameter in comparison to the larger/largest diameter crater(s) of this cluster. This moderate trend also suggests that clusters with 10 or fewer craters generally have more equally sized craters. The alignment of data points along curved trend lines seen in Figure \ref{fig:8_f_L}d is due to integer values for the number of large craters in clusters. For example, the minimum F-value is an inverse function of the total number of craters in the cluster ($1/N_c$; see Table \ref{tab:properties_description}).

\subsubsection{Differential Size-Frequency Distributions}

The number of craters per $\sqrt{2}$ diameter bin were counted for each cluster. Crater clusters with only one bin represented were not assessed further in this section, resulting in 619 crater clusters with DSFDs to analyse. The 15 crater clusters removed all had $N_c\leq3$. Identification of smaller diameter craters becomes difficult as we approach the limits of our image resolution; mapping may not always be complete. This can lead to an apparent decrease or `roll-over' of counts in bins where crater identification becomes resolution limited. It is noted that for clusters with $N_c>30$, a drop-off in crater counts can sometimes be seen in the smallest diameter bin ($D<$1.4 m; e.g. Figure \ref{fig:4_SFD}). 
For clusters with few craters this can be hard to identify and so linear regression was performed using all points (solid line in Figure \ref{fig:4_SFD}). This may however cause an underestimation of DSFD slopes ($a$ in the exponential component of the linear regression in log-log space; dashed line in Figure \ref{fig:4_SFD}). 

\begin{figure}[h]
\centering
\includegraphics[width=0.8\textwidth]{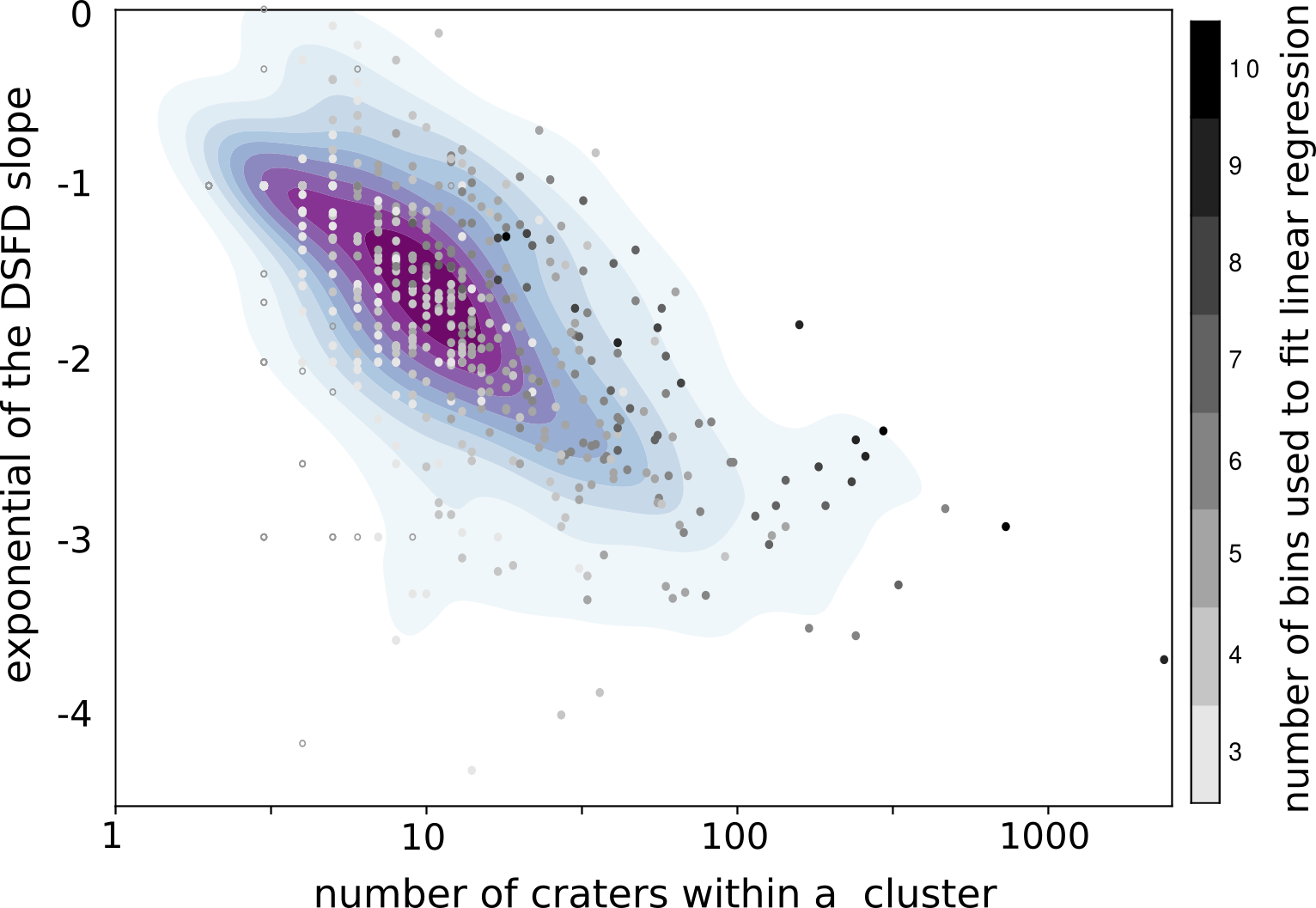}
\caption{Points show the DSFD slope for 619 crater clusters with at least two bins represented. Shading of points represents the number of bins ($>2$) used to perform the linear regression fit. Those where a fit is between two points (two bins) only are represented by open circles. Colour contouring illustrates density distribution of points only. Note that the x-axis is shown in logarithmic scale.}
\label{fig:9_SFD_all}
\end{figure}

Figure \ref{fig:9_SFD_all} shows the DSFD slopes for all 619 crater clusters analysed. It indicates there is a weak inverse correlation with the number of craters in the clusters (correlation coefficient of -0.52). Clusters with a greater number of overall craters tend to have steeper DSFD slopes.
There are 106 crater clusters with only two bins represented (open circles in Figure \ref{fig:9_SFD_all}). These craters clusters show a wide range of DSFD slopes between $-1$ and $-4$, showing high variability and are likely unreliable. Ignoring these crater clusters decreases the correlation coefficient to -0.64, strengthening the inverse correlation between number of craters and DSFD slope.

\subsection{The effect of elevation on crater cluster characteristics}
To examine the dependency of the cluster properties on elevation, we examined several cluster characteristics as a function of elevation, including the number of craters, dispersion, F-value and effective diameter.

\begin{figure}[h]
    \centering
    \includegraphics[width=\textwidth]{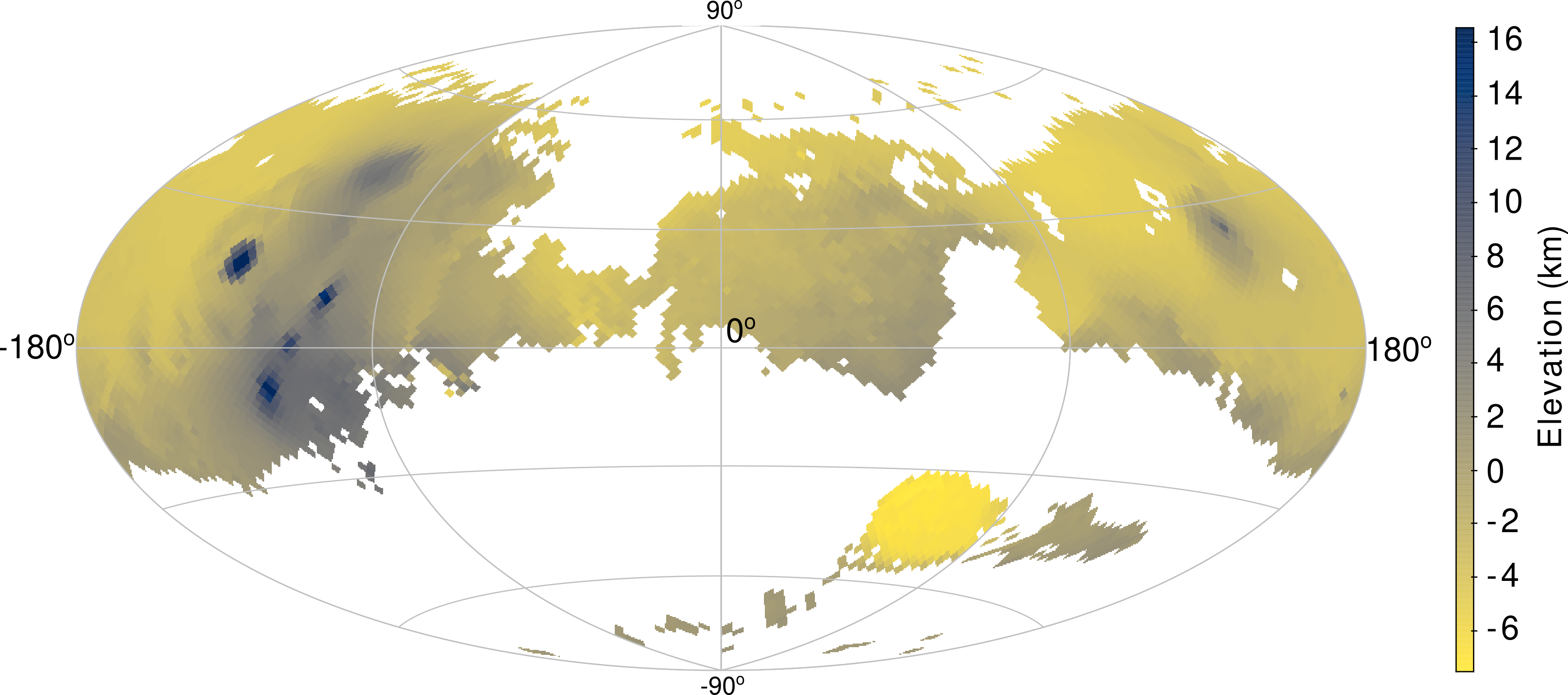}
    \caption{Aitoff projection of the martian surface, subdivided into 5th order HEALPix (the martian surface is divided into pixels of $1.83^\circ \times 1.83^\circ = 11,750\,km^2$). Colour bar shows median elevation per pixel for pixels in regions where Thermal Emission Spectrometer emissivity in the 1350 to 1400 cm$^{-1}$ wavelength $<0.95$ (hereby referred to as dusty regions).}
    \label{fig:10_aitoff_elev_dust}
\end{figure}

\begin{figure}[h]
    \centering
    \includegraphics{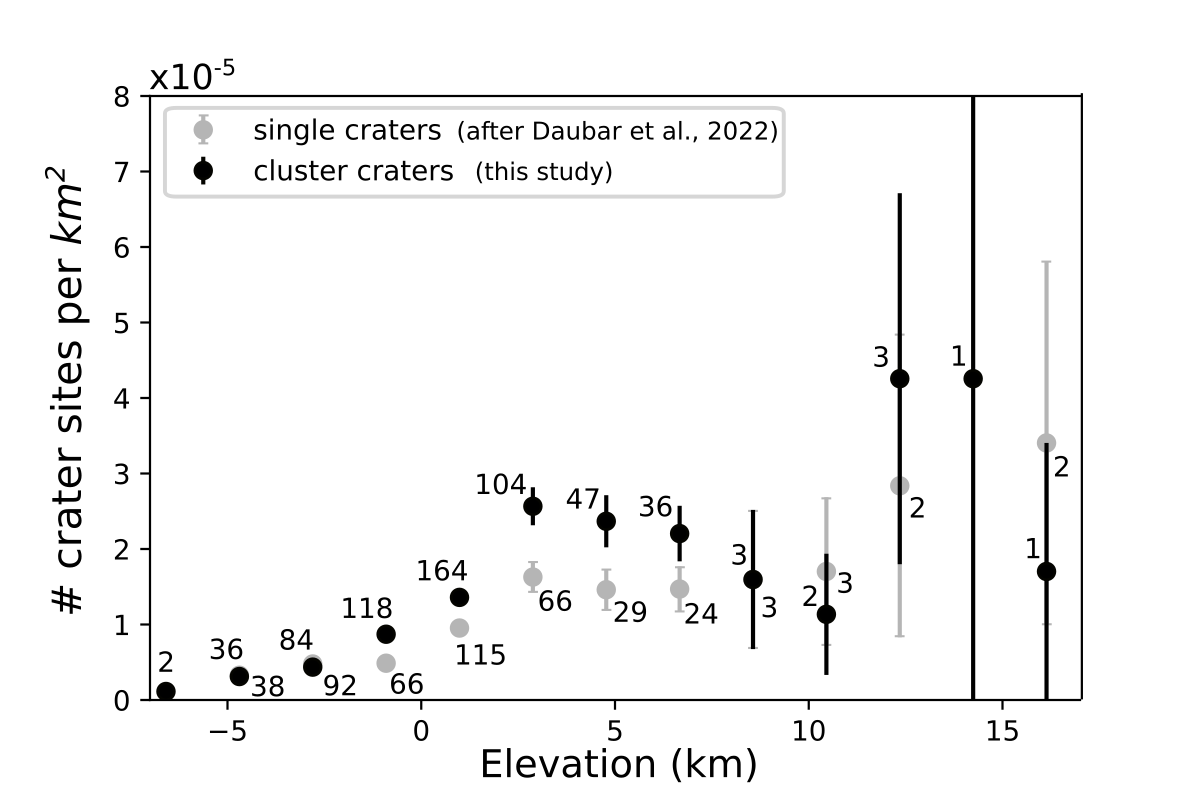}
    \caption{Number of crater cluster sites per area observed within a given elevation range. Values have been normalised to the total area covered by each elevation bin within the dusty region only of the Martian surface (for both single craters from the database of \citeA{Daubar2022} and crater clusters in this study). Counts of impact sites prior to normalisation are also shown to highlight where low sample size may influence results. However, while attempt has been made to de-bias for the dust cover, these data are biased for the observational areal coverage and frequency of images taken, as well as any form of surface erosion.}
    \label{fig:11_cl_elev_norm}
\end{figure}

Fresh craters are most easily identified in dusty regions, because in such regions impacts form large blast zones \cite{Malin2006, Daubar2019}. The area over which small craters and clusters are detected at a given elevation can vary because of varying orbital imaging frequency. Furthermore, the detection of clusters is biased by time- and space-limiting coverage of orbital imagers \cite{Daubar2022}. Here we correct for the elevation and regional dust coverage biases, but not the time component. We created a nested HEALPix map (Hierarchical Equal Area isoLatitude Pixelation of a sphere; \citeA{gorski2005healpix}) of MOLA data and TES dust cover index \cite{Ruff2002}. These pixels subdivide the martian surface into equal area bins that are identically located for each parameter we investigate. This allows us to compare the median elevation, dust index (emissivity) and crater cluster properties, as well as the number of crater clusters for each identical pixel. The median MOLA elevation per HEALPix pixel for this region is shown in Figure \ref{fig:10_aitoff_elev_dust}. The MOLA map was sourced from \citeA{Neumann2001,Neumann2003}. Dusty regions are referred to as having Thermal Emission Spectrometer emissivity of $< 0.95$ in wavelengths between
1350 to 1400 cm$^{-1}$. Within this dusty area there are 601 crater clusters (95\% of all clusters mapped). We created 15 elevation bins between the minimum and maximum values and determine the area (number of pixels) represented by each bin. We normalised crater cluster counts to give the total number of crater cluster sites per 10$^{-5}$ km$^2$ in each elevation bin (Figure \ref{fig:11_cl_elev_norm}). As such, this figure suggests a clear correlation between the number of observed cluster sites with elevation up to about 5 km altitude. Counting errors are calculated as standard errors, meaning $\sqrt{N_c}$ normalised to the area represented. Errors are large for elevations above 7.61 km due to the small area covered and small number of observed craters at these elevations. We note that this trend with elevation may be subject to further observational bias, given that orbital imagery is taken at different times and areal coverage, which may not correlate with equal representation of all elevations. 
In order to make reasonable interpretations of the relationship between elevation and meteoroid fragmentation, we perform the same analysis for the single craters in the database of \citeA{Daubar2022}, also plotted in Figure \ref{fig:11_cl_elev_norm}.

\begin{figure}[h]
    \centering
    \includegraphics[width=\textwidth]{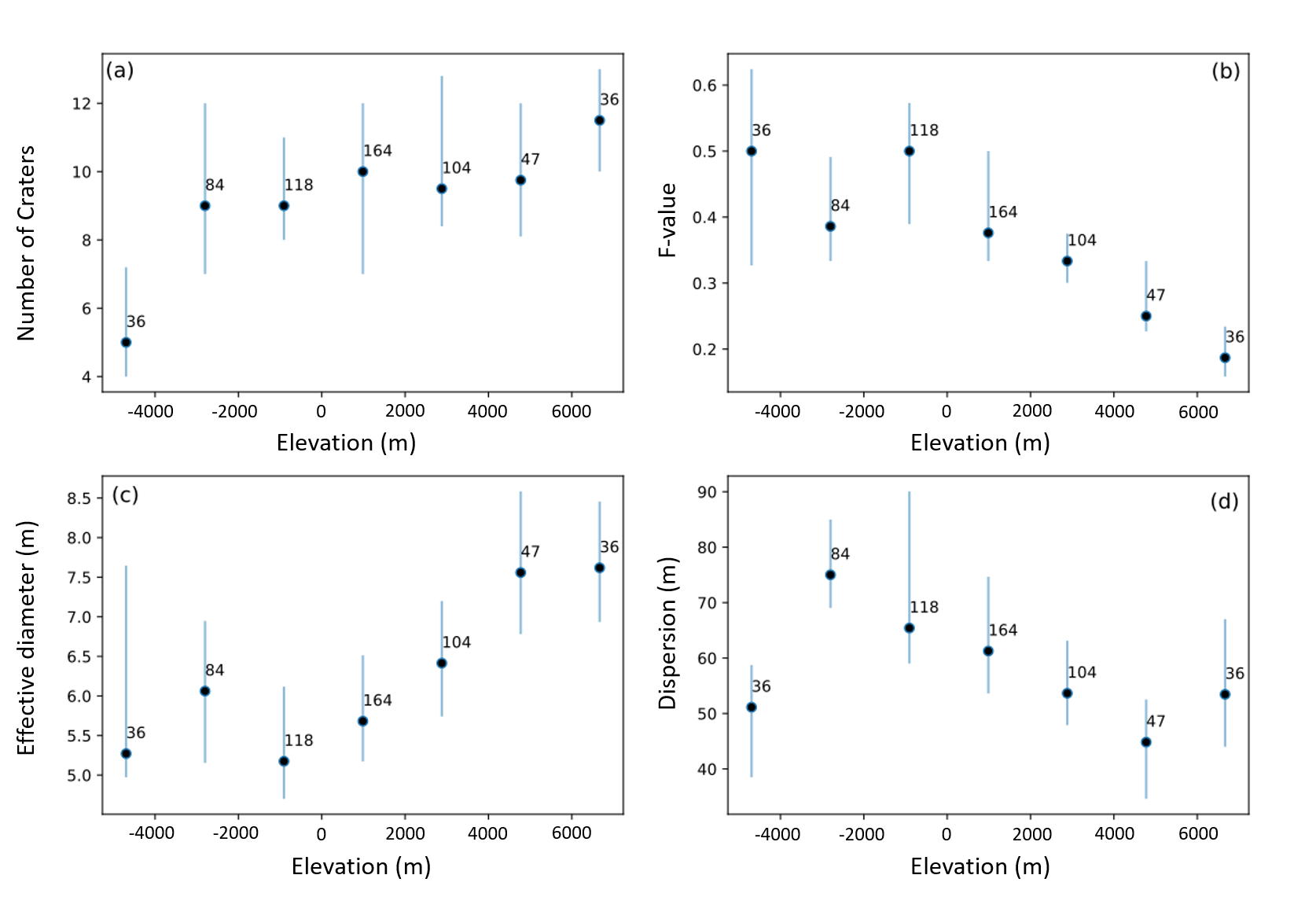}
    \caption{The variation in  crater cluster characteristics can be illustrated  by the median values of each the (a) total number of craters within a cluster; (b) F-value; (c) effective diameter; (d) dispersion of craters within a cluster. The 40th and 60th percentiles are represented by whiskers on each plot. The number of crater cluster sites falling in each bin are also given.}
    \label{fig:12_hist_elevation}
\end{figure}

Further relationships between elevation and the median value for: (a) the number of craters within a cluster, (b) F-value, (c) effective diameter, and (d) the dispersion are shown in Figure \ref{fig:12_hist_elevation}. The 40th-60th percentiles are shown as vertical error bars to give an idea of value ranges. For statistical significance, we limit this to elevation bins with more than three crater clusters, which limits the data to between -5.64 km and 7.61 km. We found that the number of craters within a cluster shows a slight increase with elevation, as does the crater cluster effective diameter. Both the F-value and dispersion show a stronger negative correlation with elevation (Figure \ref{fig:12_hist_elevation}). 

\subsection{Summary of results}\label{sec:summary}
After investigating the different parameter spaces, we find the following key trends in the data:
\begin{itemize}
    \item There is a strong correlation between the effective diameter ($D_\mathrm{eff}$) and the largest crater diameter within the cluster ($D_{max}$). The $D_{max}$ can be used as a reasonable proxy for the effective diameter of a crater cluster (which in turn is typically used as a proxy for impactor mass). See Figure \ref{fig:7_general}a.
    
    \item Crater clusters with larger effective diameters (high $D_\mathrm{eff}$) generally have high dispersion and low F-values, so consist mostly of dispersed, small diameter craters, with a low fraction of large members. See, Figures \ref{fig:7_general}d and \ref{fig:8_f_L}a.
    
    \item Crater clusters with small effective diameters (low $D_\mathrm{eff}$) do not have large numbers of individual craters and show a range of dispersions and relative crater sizes. See Figure \ref{fig:7_general} and \ref{fig:8_f_L}a.
    
    \item Crater clusters with many individual craters ($N_c>100$) generally show higher effective diameters, high dispersion, low F-values and steep DSFD slopes, so consist of many dispersed craters with a small fraction of large members. See Figures \ref{fig:7_general}b,c, \ref{fig:8_f_L}d and \ref{fig:9_SFD_all}.
    
    \item Crater clusters with low $N_c$ typically show a more even distribution of individual crater diameters, as supported by the higher F-values and shallower DSFD slopes. See Figures \ref{fig:8_f_L}d and \ref{fig:9_SFD_all}.
    
    \item With increasing elevations, crater dispersion decreases, as does the F-value, while size ($D_\mathrm{eff}$) appears to increase (see Figure \ref{fig:12_hist_elevation}). There is little correlation seen between the number of craters within a cluster and the elevation (Figure \ref{fig:12_hist_elevation}a). Overall, sites at higher elevations appear to have been more energetic (larger effective diameters from either larger impacting bodies or from increases impact speeds), though more compact, and having a small fraction of large members. 
    
    \item Elevations up to 5 km show a steady increase in the total number of crater cluster sites detected per unit area, above which they begin to decrease. See Figure \ref{fig:11_cl_elev_norm}.
\end{itemize}

\section{Discussion}

The properties of crater clusters can inform our understanding of the atmospheric effects, flight dynamics, properties of a meteoroid and any fragmentation processes it experienced. The variations with elevation and size-frequency distributions of craters in a cluster, as characterised by both the fraction of large craters in a cluster (F-value) and the slope of a crater DSFDs, may provide insight into the fracture resistance of meteoroids and the mechanisms of their break-up. 

\subsection{Variations in crater cluster properties with size and surface elevation}

Crater scaling relationships can be used to approximate the amount of momentum or energy required to form a (single) crater \cite{Teanby2011, Holsapple1993}. If an impact speed is assumed, this can provide an estimate of impactor mass. The effective diameter is calculated for a crater cluster in order to estimate the equivalent (by volume) single crater size. 
For large effective diameters, we see that dispersion is generally high, while F-values are typically low. Although the maximum number of craters that can form increases, larger clusters show a range of $N_c$ values. Small clusters on the other hand do show limited numbers of impacting fragments (low N$_c$), but a wide variation of the fraction of large craters and dispersion. 
This dispersion of craters in a cluster provides constraints on fragment flight dynamics. 
For equivalent impact velocities, this would imply larger meteoroids typically break into a few large fragments and many smaller fragments with greater spreading velocities, while small bodies are capable of a range of different sized fragments and spreading velocities. This could indicate a tendency for larger meteoroids to disrupt higher up, for fragment separation to be enhanced by bowshock interactions, or merely be an observational bias (where many smaller fragments form craters below our 1 m mapping limit). 

The relationship of these properties with elevation provide further information. 
For an impact angle of 45 degrees and a fixed meteoroid strength of 0.65 MPa, fragmentation altitudes of ordinary chondrite meteoroids 1-30 cm in diameter typically range between 10 and 25 km, although fragmentation can occur from the ground to as high as 30 km \cite{Williams2014}. The variations in bulk strengths (as suggested by \citeA{Collins2022}) also supports a large range of fragmentation altitudes. Finding crater clusters as high as 17.9\,km is therefore not unexpected. A median crater separation distance of 60 m, implies typical fragment separation speeds of 0.1--0.3\% times the meteoroid speed at break-up.  This is comparable with the typical separation velocities inferred from detailed observations of terrestrial meteoroid falls \cite<e.g.,>{Borovicka2003} and crater strewn fields \cite{Passey1980}. 

The craters mapped in this work formed on surfaces with elevations ranging approximately two atmospheric scale heights, corresponding to a factor of $\sim$9 variation in atmospheric density. For 10--100 cm-scale stony meteoroids, which likely produce the majority of clusters in our data set, this can translate to a 3--8 times variation in impact velocity at the ground and up to a five times variation in surviving mass \cite{Williams2014, Collins2022}. 

Impacts at lower elevations will hit the ground at lower speeds due to deceleration by atmospheric drag. Other atmospheric filtering effects would be expected to be observed for crater clusters at lower elevations; smaller fragments may ablate to nothing, while longer trajectories support greater fragment dispersion \cite{Hartmann2018}. 
Therefore, for otherwise identical impact scenarios, fragmented meteoroids striking Mars at lower elevations will hit the ground at lower speeds, with potentially lower surviving mass, fewer small craters, and increasing separation between the craters. This would result in a cluster with a smaller effective diameter (as seen in Figure \ref{fig:12_hist_elevation}c), and greater dispersion (as seen in Figure \ref{fig:12_hist_elevation}d). The trend in dispersion with altitude was also observed by \citeA{Daubar2019}. With smaller fragments ablating away, the fraction of large craters would be expected to increase (supported by Figure \ref{fig:12_hist_elevation}b), and may result in fewer impact sites in general (weak decrease seen in Figure \ref{fig:11_cl_elev_norm}).
Smaller effective diameters at lower elevations may also lead to clusters being less detectable in spacecraft images (Figure \ref{fig:11_cl_elev_norm}).
At higher altitudes, larger craters are expected to form, as impacts are more energetic due to higher impact speeds. The decrease in the fraction of large craters in a cluster with elevation also suggests that higher elevation terrain may be recording more impactor material from high altitude fragmentation. 

We note that \citeA{Daubar2022} observed no difference in the proportion of new impact sites that were clusters versus single craters as a function of elevation, which would indicate fragmentation and cluster formation is not more prevalent at higher elevations. This result is likely due to the variation in dynamic strength of meteoroids being much greater than the factor of $\sim9$ variation in atmospheric density \cite{Collins2022}. Ablation and deceleration, on the other hand, affect crater size for both single crater and cluster-forming impacts. As discussed above, an expected reduction in fragmentation at higher altitudes (due to lower atmospheric densities) and lower survival rate of small fragments at lower altitudes (due to ablation), we would expect to see a greater difference in the ratios single vs crater clusters with elevation, with increased numbers of single craters at lower and possibly higher altitudes. This is not clear in Figure \ref{fig:11_cl_elev_norm}, nor in the results of \citeA{Daubar2022}, however this could again be due to the difficulty in detecting craters in spacecraft images at low altitudes, and smaller areas of high altitude terrain. 

The atmospheric density and pressure on Mars has varied in the past. From isotopic measurements \cite{jakosky2001mars}, the density of Mars' atmosphere could have been $>$10x that of today \cite{fassett2011sequence}. The range of elevations for newly formed craters analysed in this study probe much of these density differences. Variations observed with elevation today could be used as a proxy for historical variations at fixed elevations \cite{chappelow_influences_2005}. Historical clusters could also potentially be used to infer palaeo-pressures of Mars' atmosphere \cite{Kite2014}.

Atmospheric filtering effects can explain the trend in detected cluster sites with elevation, as well as their dispersions and large crater fractions. These effects however, do not appear to control the number of fragments formed during atmospheric disruption of a meteoroid (observed as the total number of craters within a cluster). This must be controlled by other factors, such as the material properties of the meteoroids.

\subsection{Implications for meteoroid material properties}

The number of craters in a cluster can indicate to some extent the number of fragments produced during atmospheric breakup. The relative sizes of the individual craters in a cluster can be used to inform the sizes of these surviving fragments. Under the assumption that each crater in a cluster is formed by a single fragment, and that crater size is approximately proportional to the cube root of the fragment mass, the slope of the crater size frequency distribution in log space is approximately equal to three times the slope of the fragment mass. The slope of the size-frequency distribution of populations of fragments has been interpreted as both a measure of the degree of comminution \cite{Hartmann1969} and the resistance to fracturing of the fragmented material \cite{Turcotte1986}. 

Meteoroids with a large number of fragments surviving to impact (high $N_c$) show steep DSFD slopes (more negative exponential; Figure \ref{fig:9_SFD_all}). Clusters with steeper DSFD slopes could indicate early breakup with many surviving fragments, or that the material is much weaker or greatly fractured. 
\citeA{Hartmann1969} showed that fragment populations with a shallow DSFD slope are typical outcomes of break-up by a single or small number of fragmentation events, such as sledgehammer blows or low velocity impacts, whereas populations with steeper slopes were produced by repeated low-energy fragmentation (comminution) or high-energy explosive fragmentation (for example, by hypervelocity impact). On the other hand, \citeA{Turcotte1986} showed that the power-law slope of fragment size frequency distribution can also be interpreted as a measure of how resistant a material is to a given fragmentation process. In this interpretation, a shallow slope implies a lower resistance to fracturing or higher degree of fragility, whereas a steeper slope implies a higher resistance to fracturing or more resilient material response. 

The large variation in DSFD slope among martian crater clusters (Fig. \ref{fig:9_SFD_all}) can therefore be interpreted in one of three ways. One interpretation is a diversity in dynamic fragmentation style, which might be related to the magnitude or rate of applied dynamic stress. In this case, failure may range from a single or small number of relatively low-energy fragmentation events, perhaps resulting from a relatively low-velocity or shallow angle meteoroid trajectory, to either a catastrophic high-energy disruption (e.g., high-velocity and/or high-angle trajectory) or a sustained progression of fragmentation events that “grind” the meteoroid down to its final population of fragments. A second interpretation is that the range of DSFD slopes reflects a wide range of meteoroid fragility. Shallow slopes are produced by meteoroids with low fracture resistance (i.e., more internal weaknesses), whereas steeper slopes are produced by meteoroids more resilient to fracturing. We note that fracture resistance may not be the same thing as dynamic strength---i.e., a meteoroid may have a relatively high fragility and still begin to fragment at a relatively high dynamic stress if, for example, the meteoroid represents a sintered breccia. The third interpretation is that rather than probing the fragmentation process directly, the observed range of DSFD slope is merely a reflection of the diversity in the underlying internal particle size-frequency distributions of the population of meteoroids entering Mars’ atmosphere. The best way of discriminating between these would be to observe the fireball phenomena itself. Until a meteor or fireball is observed on Mars, future work using Earth analogues from dedicated camera networks \cite<e.g.>[]{Devillepoix2020, borovivcka2022data, vida2021global} may aid in deducing the more plausible scenario.

\section{Conclusion}

The measured distributions of crater cluster properties on Mars provides insight into the meteoroid fragmentation and separation process and provides constraints for meteoroid atmospheric flight and fragmentation models.
This study describes the properties of 634 crater clusters recently formed on Mars, combining 557 newly mapped crater clusters with 77 previously investigated by \citeA{Daubar2019}. 

When interrogating this much larger data set, we find a diversity in the properties of crater clusters. 
There are general trends that support atmospheric filtering effects on impacting material: crater clusters at lower elevations have smaller effective diameters and are generally more dispersed, reflecting increased ablation, lower impact speeds and greater lateral spreading of fragments. The trends are not as pronounced as we might expect, and there does not appear to be a clear relationship between elevation and single vs crater cluster formation. This implies that meteoroids impacting as single bodies vs those that fragment in the atmosphere do not appear to favour any specific pressure regimes. This indicates a variation in the densities and strengths of the impacting population, and any trajectory models need to account for this diversity. At this stage, the upper limit of meteoroid size able to form a cluster crater is not well constrained; we see clusters up to nearly the largest diameters that we see in single craters \cite{Daubar2022}.

Although atmospheric filtering accounts for differences in size and dispersion of crater clusters, the number of craters within a cluster (N$_c$) shows little trend with elevation. The variation in N$_c$ is therefore more likely related to the material properties of the impactor, such as its bulk strength. Crater clusters with a large number of craters are more widely dispersed and are typically dominated by a few larger craters. This would be expected for body fragmentation high above the ground, and overall weaker bulk strength.   

Mars' crater cluster population samples elevations that represent a similar range of atmospheric densities to those that are expected to have existed in Mars' past. The weak trends we observe in the  properties of crater clusters due to atmospheric filtering indicate that cratering mechanisms may have been similar in the past.  

\acknowledgments
TN is fully supported by the Australian Research Council on DP180100661. KM is fully supported by the Australian Research Council on DP180100661 and FT210100063. EKS is fully supported by the Australian Research Council on DP200102073. IJD is supported by NASA Solar System Workings grant 80NSSC20K0789. GSC was funded by UK Space Agency Grant grant ST/T002026/1. 

\section*{Open Research}
\noindent Data from the crater cluster mapping and calculated parameters used for analysis can be found in the Zenodo archive \cite{zenodo}. 

\bibliography{references}

\end{document}